\documentclass[aip,amsmath,amssymb,preprint]{revtex4-2}

\usepackage{graphicx}% Include figure files
\usepackage{dcolumn}% Align table columns on decimal point
\usepackage{bm}% bold math
\usepackage{hyperref}% add hypertext capabilities
\usepackage{amsmath}
\usepackage{xcolor, soul}
\sethlcolor{yellow}
\usepackage{comment}
\usepackage{gensymb}
\usepackage{siunitx}
\usepackage{natmove}
\usepackage{tabularx}
\usepackage{float}

\newcommand{\mose}{MoSe$_2$}

\newcommand{\tise}{TiSe$_2$}

\newcommand{\xt}{X$^-$}
\newcommand{\xo}{X$^0$}

\newcommand{\cmi}{cm$^{-1}$}
\DeclareUnicodeCharacter{2212}{-}
\newcommand{\moire}{moir\'{e}}

\newcommand*{\citen}[1]{%
  \begingroup
    \romannumeral-`\x % remove space at the beginning of \setcitestyle
    \setcitestyle{numbers}%
    \cite{#1}%
  \endgroup   
}

\def\introfig{
        \begin{figure*}[tbp]
            \centering
            \includegraphics[width=6.67in]{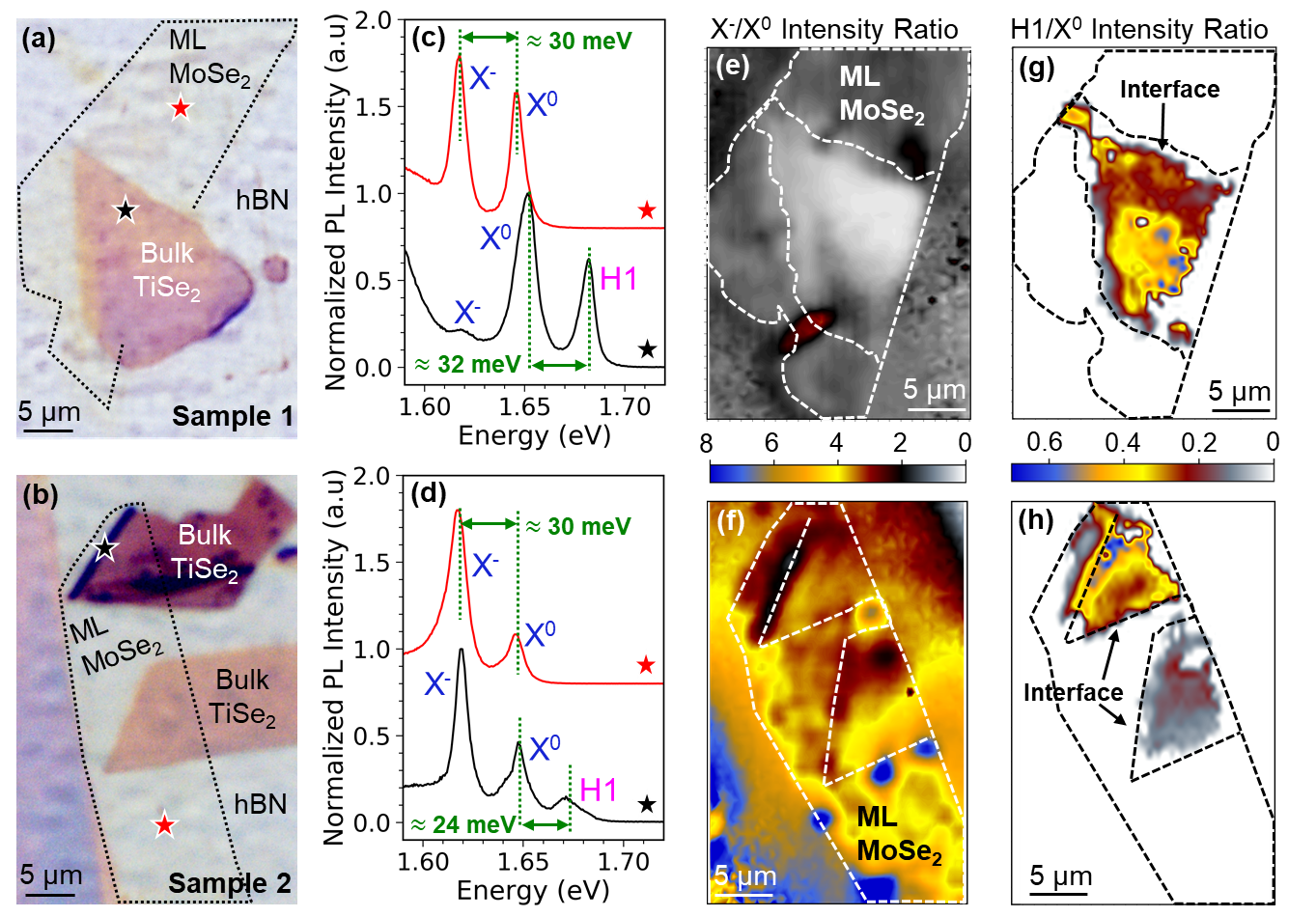}
            \caption{Optical microscope images of \tise{} - \mose{} vdW heterostructures (a) Sample 1 and (b) Sample 2. The black dotted line outlines ML-\mose{} layer. Low-temperature (5 K) photoluminescence (PL) spectra taken on (black) and off (red) the interface for (c) Sample 1 and (d) Sample 2. These locations are denoted by red and black stars in panels (a) and (b), respectively. In addition to the \xo{} and \xt{} emission observed in ML-\mose, the H1 PL peak appears at $\approx$ 1.68 eV on the interface. Spatially mapped \xt{}/\xo{} integrated intensity ratio across the interface showing variations in the \xt{} intensity for (e) Sample 1 and (f) Sample 2. Spatially mapped H1/\xo{} integrated intensity ratio across the interface showing changes in H1 intensity in (g) Sample 1 and (h) Sample 2. The ratio \xt{}/\xo{} correlates with changes in H1/\xo{} intensity at the junction.} 
            \label{introfig}
        \end{figure*}
        }

\def\temppower{
        \begin{figure*}[tbp]
            \centering
            \includegraphics[width=6in]{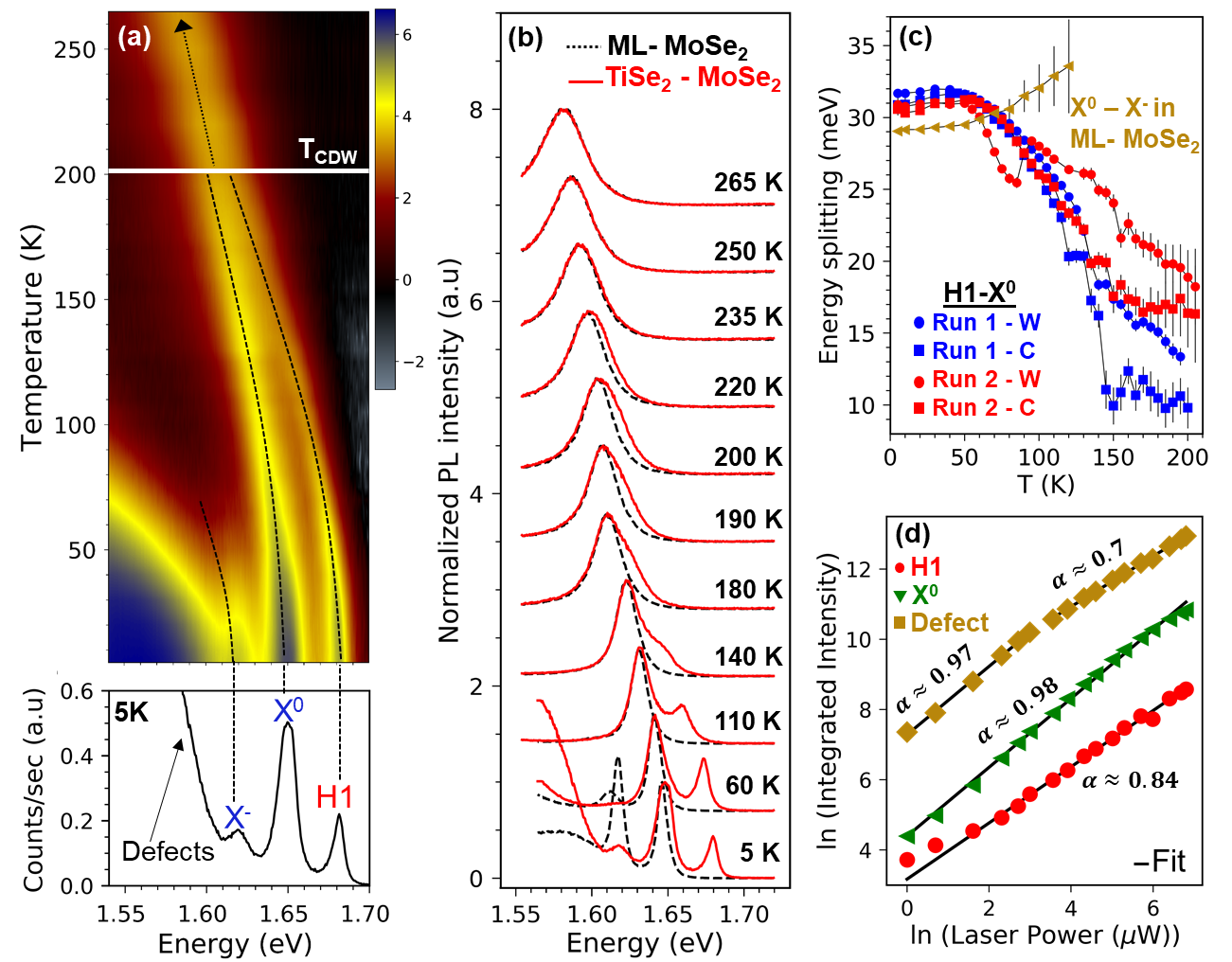}
            \caption{(a) Temperature-dependent PL map on the \tise{} - \mose{} interface of Sample 1. The dashed lines are guides to the eye and the 5 K PL spectrum beneath the map labels the optical transitions. $T_{CDW}$ for bulk \tise{} is indicated by the horizontal white line. (b) Lineshape analysis of PL spectra taken on (red) and off (black) the \tise{} - \mose{} interface. The interface spectra have been shifted to align the \xo{} emission energies between the two curves. (c) Energy separation between H1 and \xo{} versus temperature for two separate runs (red, blue), each consisting of a warming (W, circles) and cooling (C, squares) curve. 1-$\sigma$ error bars from fits to the PL spectra are included. The \xo{}-\xt{} is included for ML-\mose{} to illustrate the difference from H1. (d) Log-log plot of the PL integrated intensity \textit{versus} excitation laser power. The solid black lines are power-law fits to the data.}
           \label{temppower}
       \end{figure*}
        }

\def\ramandata{
        \begin{figure}[tbh]
            \centering
            \includegraphics[width=3.4in]{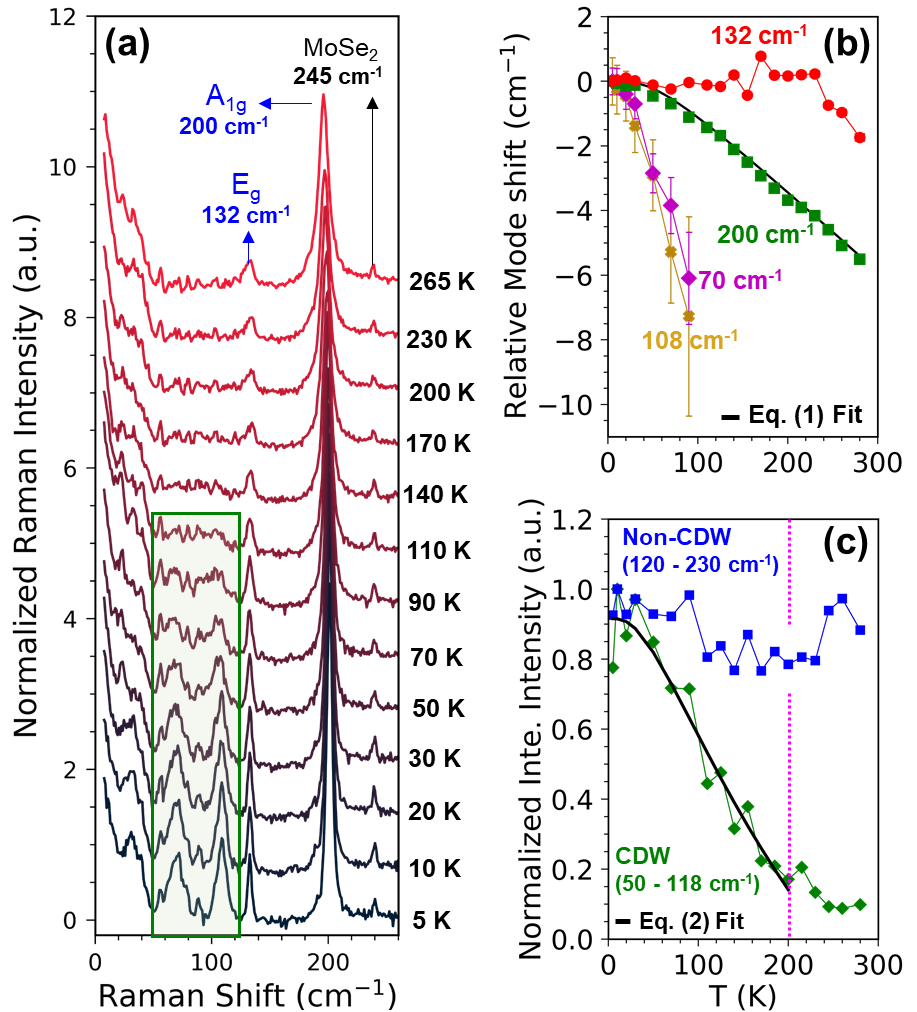}
            \caption{(a) Temperature-dependent Raman spectra taken on the \tise{} - \mose{} interface in Sample 1. (b) Shift in the observed \tise{} Raman modes relative to the 5 K value. 1-$\sigma$ error bars from spectral fitting included. The solid black line is fit to Eq. (\ref{balkanksi}) that accounts for optical phonon decay processes (c) Normalized integrated Raman intensity of the CDW modes (green diamonds) and lattice modes (blue squares) versus temperature. The dotted magenta line indicates $T_{CDW}$. The solid black line is a fit to Eq. (\ref{eq:1}).}
           \label{ramandata}
       \end{figure}
        }

\def\grand{
        \begin{figure*}[!h]
            \centering
            \includegraphics[width=6.67in]{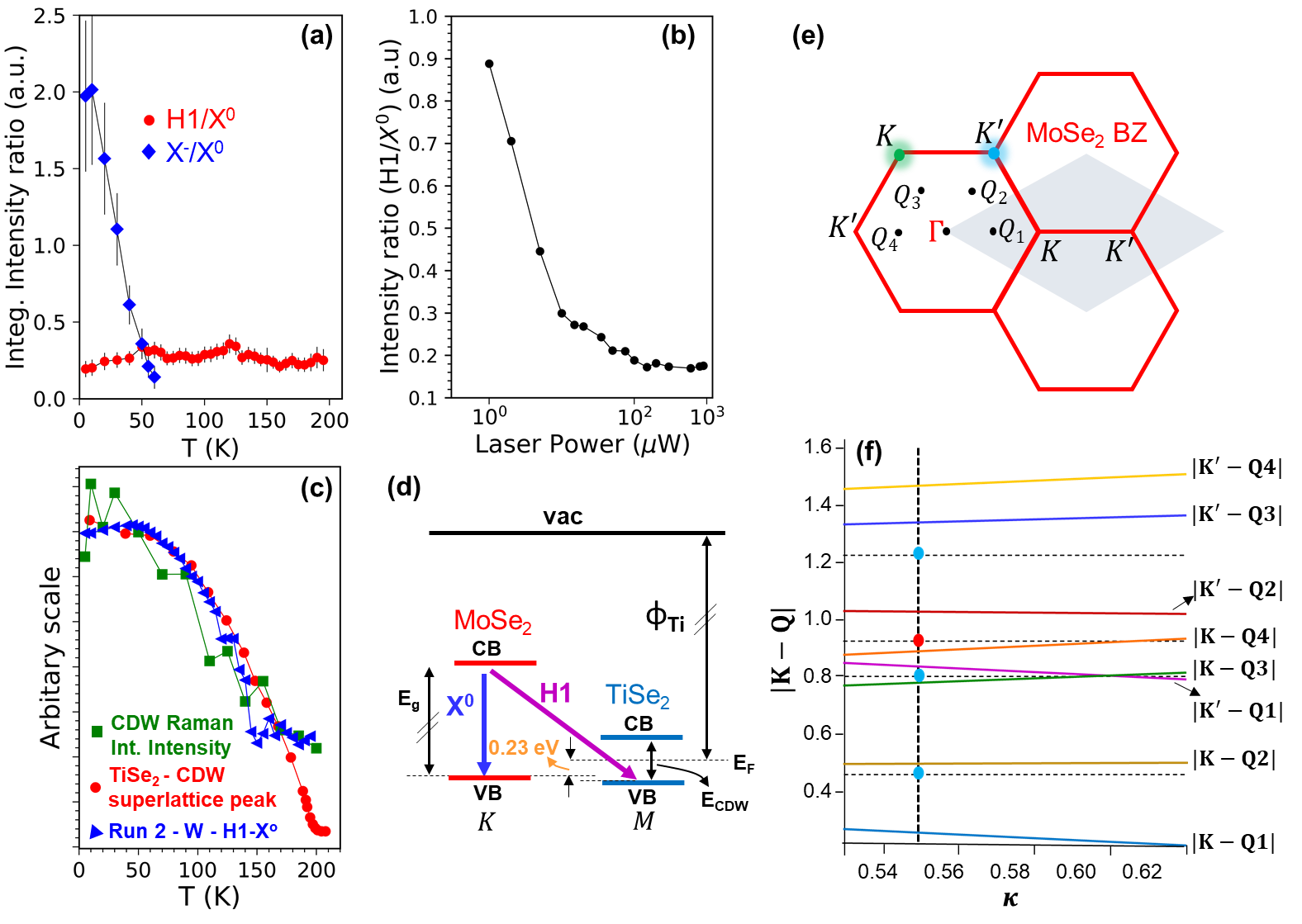}
            \caption{(a) PL integrated intensity ratios H1/\xo{} (red circles) and \xt{}/\xo{} (blue diamonds) versus temperature. (b) H1/\xo{} integrated intensity ratio versus the laser power (lin-log scale). (c) Overlay of the energy separation between H1 and \xo{} for Sample 1, the integrated Raman intensity of the CDW modes and the \tise{}-CDW superlattice peak extracted from neutron scattering data in Ref. \protect\citen{DiSalvo1976}, all as a function of temperature. (d) A schematic of a  type-II band alignment at the \tise{} - \mose{} interface leading to an interlayer exciton. (e) Brillouin zones for the \mose{} lattice and the corresponding r.l. unit cell (gray). (f) Lengths of the scattering vectors $\mid \mathbf{K-Q} \mid$ in units of the \mose{} reciprocal lattice (r.l.) vectors (colored solid lines) compared to the length of the reciprocal lattice vectors of the undistorted (red dot) and CDW (blue dots) \tise{}. The horizontal dashed lines are guides to the eye. The vertical dashed line is drawn to pass through $\kappa = 0.55$, to facilitate the comparison with the magnitude of the \mose{} wavevectors $\mathbf{|K-Q|}$ and the smallest r.l. vector (dots) in \tise{}.}
           \label{grand}
       \end{figure*}
}

\begin{document}
\title{Charge Density Wave Activated Excitons in \tise{} - \mose{} Heterostructures}

\begin{abstract}
Layered materials enable the assembly of a new class of heterostructures where lattice-matching is no longer a requirement. Interfaces in these heterostructures therefore become a fertile ground for unexplored physics as dissimilar phenomena can be coupled via proximity effects. In this article we identify an unexpected photoluminescence (PL) peak when \mose{} interacts with \tise{}. A series of temperature-dependent and spatially-resolved PL measurements reveal this peak is unique to the \tise{} - \mose{} interface, higher in energy compared to the neutral exciton, and exhibits exciton-like characteristics. The feature disappears at the \tise{} charge density wave transition, suggesting that the density wave plays an important role in the formation of this new exciton. We present several plausible scenarios regarding the origin of this peak that individually capture some aspects of our observations, but cannot fully explain this feature. These results therefore represent a fresh challenge for the theoretical community and provide a fascinating way to engineer excitons through interactions with charge density waves.
\end{abstract}

\author{Jaydeep Joshi}
\affiliation{Department of Physics and Astronomy, George Mason University, Fairfax, Virginia 22030, United States}
\affiliation{Quantum Science and Engineering Center, George Mason University, Fairfax, Virginia 22030, United States}

\author{Benedikt Scharf}
\affiliation{Institute for Theoretical Physics and Astrophysics and Würzburg-Dresden Cluster of Excellence ct.qmats, University of Würzburg, Am Hubland, 97074, Würzburg, Germany }

\author{Igor Mazin}
\affiliation{Department of Physics and Astronomy, George Mason University, Fairfax, Virginia 22030, United States}
\affiliation{Quantum Science and Engineering Center, George Mason University, Fairfax, Virginia 22030, United States}

\author{Sergiy Krylyuk}
\affiliation{Materials Science and Engineering Division, National Institute of Standards and Technology, Gaithersburg, Maryland 20899, United States}

\author{Daniel J. Campbell}
\affiliation{Maryland Quantum Materials Center, Department of Physics, University of Maryland, College Park, Maryland 20742, United States}

\author{Johnpierre Paglione}
\affiliation{Maryland Quantum Materials Center, Department of Physics, University of Maryland, College Park, Maryland 20742, United States}
\affiliation{Canadian  Institute  for  Advanced  Research,  Toronto,  Ontario  M5G  1Z8,  Canada}

\author{Albert Davydov}
\affiliation{Materials Science and Engineering Division, National Institute of Standards and Technology, Gaithersburg, Maryland 20899, United States}
\affiliation{Quantum Science and Engineering Center, George Mason University, Fairfax, Virginia 22030, United States}
\affiliation{Maryland Quantum Materials Center, Department of Physics, University of Maryland, College Park, Maryland 20742, United States}

\author{Igor \v{Z}uti\'{c}}
\affiliation{Department of Physics, University at Buffalo, Buffalo, New York 14260, United States}

\author{Patrick M. Vora*}
\affiliation{Department of Physics and Astronomy, George Mason University, Fairfax, Virginia 22030, United States}
\affiliation{Quantum Science and Engineering Center, George Mason University, Fairfax, Virginia 22030, United States}
\email{pvora@gmu.edu}

\maketitle

\section{Introduction}
Two-dimensional (2D) material interfaces in van der Waals (vdW) heterostructures provide a fascinating playground to explore proximity effects. \cite{Zutic2019}. The relaxation of lattice constraints on heterostructure assembly allows for the arbitrary stacking of 2D materials.\cite{Geim2013} These interfaces may, in some cases, support emergent states absent from the parent compounds, with superconductivity in twisted bilayer graphene and \moire{} excitons in transition metal dichalcogenides (TMDs) serving as remarkable examples.\cite{Bistritzer2011,Cao2018,Tran2019,Yu2017} Many studies of vdW heterostructures incorporate semiconducting TMDs as an active component. This commonality is due to the availability of high-quality samples, well-established exfoliation procedures,\cite{Frisenda2018a,Onodera2019} and the existence of tightly-bound 2D excitons.\cite{Wang2017a}

The zoology of excitons in monolayer (ML) semiconductors is vast: neutral (\xo{}) and charged excitons or trions (\xt{} or X$^+$) \cite{Wang2017a,Mueller2018,Ross2013,Li2020a}, neutral and charged biexcitons \cite{Li2018,Ye2018a,Chen2018b,Barbone2018} and dark exciton states \cite{Robert2020,Lu_2019,Malic2017,Zhang2017a} have all been observed and exhaustively studied in semiconducting TMDs. The 2D nature of TMD excitons also renders them highly sensitive to the local dielectric environment \cite{Hsu2019,Florian2018,Borghardt2017,Stier2016,Raja2017,Raja2019} allowing for a remote, contact-free probe of interface characteristics in vdW heterostructures. For instance, semiconductor hetero-bilayer and homo-bilayer heterostructures exhibit new PL emission peaks from interlayer excitons and splitting of exciton peaks due to the \moire{} potential.\cite{Jiang2021,Hanbicki2018,Miller2017,Tran2019,Zhang2018b} Proximity effects between 2D magnets and semiconductors lead to large valley splittings,\cite{Norden2019,Ciorciaro2020} and magnetic manipulation of exciton PL energy, intensity, and selection rules.\cite{Scharf2017a,Zhong2017,Zutic2019,Xu2020} Exotic correlated insulating states such as Wigner crystals and Mott insulators in twisted TMD semiconductor heterostructures are also observable in PL spectra.\cite{Zhou2020b,Liu2020a,miao_strong_2021} However, there have been no studies exploring the impact of similar electron correlated phases such as charge density waves (CDWs) on PL in vdW heterostructures.

Here we investigate optical signatures of interlayer coupling between the semiconductor \mose{} and the putative excitonic insulator 1T-\tise{},\cite{Campbell2019,Rossnagel2002,Pillo2000,Cercellier2007,Monney2009} which hosts a commensurate 2 $\times$ 2 $\times$ 2 CDW state below 200 K. We find that the CDW alters the manifold of optically-active excitons at the \tise{} - \mose{} interface, which results in a new PL peak above \xo{}. This feature, referred to as H1, appears in the \mose{} PL spectrum with a linewidth comparable to \xo{}. While lower-energy PL sidebands are relatively common in TMDs due to phonon replicas and exciton localization \cite{He2020,Liu2020b,Brem2020a,Li2019c,Li2020}, these observations are the first such detection of a higher-energy PL sideband. Detailed temperature-, power-, and spatially-resolved PL measurements on multiple heterostructures demonstrate that H1 has an origin consistent with a native exciton state rather than a localized exciton or defect state. However, H1 disappears at the \tise{} CDW temperature which suggests these two phenomena are closely linked. We have identified multiple plausible scenarios and discuss them in detail, although none are able to explain all aspects of our observations. Interactions between excitons and CDWs provide a fresh challenge to the theoretical community and a novel method for engineering excitons in 2D materials.

\introfig{}

\section{Results}

\subsection{Optical characterization of vdW heterostructures}

Optical microscope images of two \tise{} - \mose{} vdW heterostructures termed Sample 1 and Sample 2, respectively, are presented in Figs. \ref{introfig}(a) and \ref{introfig}(b). These samples are assembled by a modified viscoelastic method\cite{Castellanos-Gomez2014} that incorporates atomic force microscope (AFM) cleaning.\cite{Rosenberger2018} The black dashed line outlines the ML-\mose{} flake in each sample. Representative low-temperature (5 K) PL spectra on (black) and off (red) the interface are shown in Figs. \ref{introfig}(c) and \ref{introfig}(d) for the two samples. Emission from the \mose{} \xt{}($\approx$ 1.62 eV) and \xo{}($\approx$ 1.65 eV) states agrees with prior observations in both energy and linewidth \cite{Lu_2019,Ajayi2017,Wierzbowski2017,Jadczak2017a,Ross2013}. The interface PL spectra contains a previously unobserved feature, referred to as H1, at $\approx$ 1.68 eV. H1 is comparable to \xo{} in both intensity and linewidth for Sample 1, while being weaker and broader in Sample 2. 

The 5 K PL spectra also show evidence of an anti-correlation between H1 and \xt{}. Spatially resolved PL maps of the interface allow us to explore this behavior further and connect it to interface quality by examining the integrated intensity ratios \xt{}/\xo{} (Figs. \ref{introfig}(e) and \ref{introfig}(f)) and H1/\xo{} (Figs. \ref{introfig}(g) and \ref{introfig}(h)). H1 and \xo{} + \xt{} integrated intensities are plotted separately in Supporting Information Fig. S1. For Sample 1, the ratio of \xt{}/\xo{} (Fig. \ref{introfig}(e)) varies between 1-2 over most of the \mose{} flake with a notable jump at the crack on the bottom left quadrant of the map. On the heterostructure itself, this ratio plummets to well below 1 indicating the absence of free charges that can participate in trion formation. The connection of the \xt{}/\xo{} ratio to charge transfer is well established in numerous reports\cite{Ross2013,Li2018,Liu2019e,Li2020a} and, since the transfer efficiency is exponentially dependent on distance, can be used as a proxy for interlayer spacing. \mose{} tends to be $n$-type as-exfoliated and \tise{} band alignment suggests it will act as an electron acceptor.\cite{Zhang2016a} Therefore, the near absence of \xt{} emission on the Sample 1 overlap region suggests good coupling between the \tise{} - \mose{} flakes. For Sample 2 the \xt{}/\xo{} ratio (Fig. \ref{introfig}(f)) is larger on the \mose{} flake, varying between 3-6, which may originate from unintentional doping during the heterostructure fabrication process. On the \tise{} - \mose{} overlap we observe a reduction in \xt{} intensity co-localized with H1, but smaller than in Sample 1. From this we suggest that interlayer coupling is weaker in Sample 2 which could be due to contaminants or partial oxidation of the \tise{} flake. The remaining analyses will therefore focus on Sample 1, unless otherwise noted.

\subsection{Temperature and power-dependence of H1}

\temppower{}

In this section we discuss temperature- and power-dependent PL measurements on and off the \tise{} - \mose{} interface. Figure \ref{temppower}(a) shows a temperature-dependent PL intensity map from 5 K- 265 K taken on the interface in Sample 1. A similar data set for Sample 2 is included in the Supplementary Information (Fig. S2). Here, the 5 K PL emission spectrum is the same as in Fig. \ref{introfig}(c) with prominent, sharp emission from \xo{} and H1, heavily reduced \xt{} emission  and a broad feature originating from defects. With increasing temperature, PL from defect excitons and \xt{} decreases and becomes unobservable for $T>$ 70 K in agreement with prior studies.\cite{Sharma2020,Verhagen2020,Tangi2017,Tongay2013a} Both \xo{} and H1 are visible at elevated temperatures, but are difficult to distinguish above $\approx$ 190 K. We obtain a better understanding through analysis of the PL lineshape for ML-\mose{} and \tise{} - \mose{}. Figure \ref{temppower}(b) compares PL spectra at selected temperatures taken on (red curve) and off (black curve) the \tise{} - \mose{} interface. The presence of the \tise{} capping layer causes \xo{} to blueshift due to the different dielectric constant.\cite{Raja2017,Raja2019} To facilitate comparison, we eliminate this shift by adjusting the energy axis of the \tise{} - \mose{} spectrum so that the \xo{} PL peaks overlap. The energy shift amounts to $\sim$ 1-3 meV across the entire temperature range. H1 is visible as a weak shoulder of \xo{} between 190 K and 220 K. Above these temperatures, the PL lineshape on and off the interface is identical, indicating the driving mechanism behind H1 has dissipated. We also fit each PL spectrum to a sum of Lorentzian functions and extract the temperature-dependent peak parameters for H1, \xo{}, and \xt{}. The energy splitting, H1--\xo{}, is shown in Fig. \ref{temppower}(c) over two cooling and warming runs that extend up to 200 K. In both cases the energy separation between H1 and \xo{} decreases with temperature until 200 K after which it is difficult to obtain a reliable fit.

The temperature window of 190 K -- 220 K is apparently crucial to H1 and is known to be important for \tise{}. Bulk 1T-\tise\ undergoes a 2 $\times$ 2 $\times$ 2 commensurate-CDW transition in the range of $T_{CDW}\approx$ 200 K - 210 K, as observed in a variety of optical and electronic measurements \cite{Campbell2019,Sugawara2016,DiSalvo1976,Campbell2019,Chen2016b}. The CDW transition opens a band gap at the \tise{} $M$ point in the Brillouin zone (BZ) with an associated order parameter well described by Bardeen-Cooper-Schrieffer (BCS) model \cite{Chen2016b}. The observed correlation of H1 with $T_{CDW}$ suggests a close relationship between H1 in \mose{} and the CDW in \tise{}.

Power-dependent PL measurements provide further insight into the nature of H1, as presented for Sample 1 in Fig. \ref{temppower}(d) and for Sample 2 in the Supporting Information (Fig. S2). PL intensity generally scales with power as $I_{PL} \propto P^{\alpha}$, where $I_{PL}$ is the integrated PL intensity and $P$ is the excitation power. The exponent ${\alpha}$ $\approx$ 1 for free excitons and $>1$ for multiexcitons.\cite{Barbone2018} Localized exciton states exhibit a more complicated behavior. At low powers $\alpha\sim 1$ but then becomes sublinear as the localized states are saturated. \cite{Tongay2013a} These behaviors are observed in Fig. \ref{temppower}(d) where we plot the natural logarithms of $I_{PL}$ and $P$. Linear fits to this data allow for the extraction of $\alpha$. We find that both \xo{} and H1 have an $\alpha \approx 1$, with the value for H1 being somewhat lower, suggesting free exciton characteristics. As expected, the defect band first exhibits $\alpha \approx 1$ at low powers and then shows signs of saturation with $\alpha \approx 0.7$. While $\alpha$ for H1 is lower than would be expected for a free exciton, the absence of saturation is more consistent with this interpretation.

\subsection{CDW phase at the \tise{} - \mose{} interface}
\ramandata{}

Raman spectroscopy can probe the square of the CDW order parameter directly, since the intensity of the symmetry-forbidden modes appears in the second order of the ionic displacements of the high-symmetry positions. Raman measurements performed at the \tise{}--\mose{} interface in the range of 5 - 265 K are shown in Fig. \ref{ramandata}(a). The 5 K Raman spectra show the CDW modes at $E_g^{CDW}$ (70 \cmi{}) and $A_g^{CDW}$ (108 \cmi{}), as well as normal \tise{} lattice modes at 132 \cmi{} and 200 \cmi{} of $E_g$ and $A_{1g}$ symmetry, respectively.\cite{Duong2017,Snow2003a} The \mose{} $A_{1g}$ mode is also visible at 245 \cmi{} and is related to out-of-plane vibrations.\cite{Tonndorf2013} Figure \ref{ramandata}(b) plots the temperature-dependent shifts of all \tise{} modes relative to their frequency at 5 K. The 135 \cmi{} mode is largely unaffected by temperature changes, blueshifting slightly as temperature is lowered and then stabilizing. The 200 \cmi{} mode is insensitive to the CDW transition and its anharmonicity can be understood by a combination of optical phonon decay and temperature-dependent changes in the lattice constants.\cite{Balkanski1983}. This is described by Eq. (\ref{balkanksi}) (solid black line in Fig. \ref{ramandata}(b)). 
\begin{equation}
    \Delta(\omega(\mathrm{0}),T) = \omega_0 + A\Big(1+\frac{2}{e^x-1}\Big)
    \label{balkanksi}
\end{equation}
Here, $x= \hbar\omega_B / 2k_B T$, $\omega_0$ is the 0 K harmonic frequency and $A$ represents the anharmonic contributions to the frequency of 200 \cmi{} optical mode as it decays into two acoustic phonons. The obtained value for A = -1.85 \cmi{} is within the ballpark of similar phonon anharmonicity studies done on TMDs.\cite{Joshi2016,Su2014a}

The CDW modes at 70 \cmi{} and 109 \cmi{} redshift and broaden with increasing temperature. These modes are unresolved above $T$ = 100 K, which is well below $T_{CDW}$. This behavior is commonly attributed to quantum fluctuations of the density wave.\cite{Gruner1994} Integrating the Raman intensity over the spectral range encompassing the CDW modes allows us to monitor the CDW up to the transition temperature (Fig. \ref{ramandata}(c)). Thermal melting of the CDW is expected to follow a temperature dependence consistent with the BCS treatment where the order parameter $\Delta (T)$ can be given as\cite{wolfle_theory_1976}
\begin{equation}
    \frac{\Delta^2(T)}{\Delta^2(0)} \propto \tanh^2\left(\alpha_{BCS}\sqrt{1-\frac{T}{T_{CDW}}}\right),
    \label{eq:1}
\end{equation}
This model fits the integrated intensity data well up to $T_{CDW}$ as shown in Fig. \ref{ramandata}(c).

\section{Discussion}

Investigations into the properties of 2D excitons in TMDs have covered remarkable ground over the past decade.\cite{Mak2010,shree_guide_2021, Wang2017a, mueller_exciton_2018} Despite these remarkably comprehensive studies, no observations of vdW heterostructures have shown PL satellites above \xo{}. Therefore, we explore here three possible mechanisms that could be responsible for H1.

Before proceeding with the analysis of potential microscopic interpretations of the new peak, we summarize key experimental observations. First, H1 lies 25--32 meV above \xo{} with an intensity that differs between samples (Fig. \ref{introfig}). Second, within a single sample the relative intensity of H1 to \xo{} is temperature independent (Fig. \ref{grand}(a)). Third, the energy difference between H1 and \xo{} follows the CDW order parameter $\Delta^2$. We illustrate this in Fig. \ref{grand}(c) by overlaying neutron scattering data from Ref.\,\citen{DiSalvo1976}, the energy separation H1--\xo{}, and the integrated intensity of the \tise{}-CDW Raman features. Fourth, the intensity ratio H1/\xo{} decreases with increasing laser power (Fig. \ref{grand}(b)). And lastly, H1 and \xt{} are spatially anti-correlated (Fig. \ref{introfig}).

\subsection{Mechanism I: Activation of Forbidden/Dark Excitons}

Semiconducting TMDs host numerous dark exciton states where optical recombination is forbidden by momentum conservation or symmetry. Density functional theory (DFT) calculations of \mose{} suggest a finite-momentum dark exciton lies 30 meV above \xo{}\cite{Deilmann2019}. This indirect exciton is formed from an electron residing at the $\mathit{Q}$($\mathit{Q^\prime}$) valley and a hole at the $\mathit{K}$($\mathit{K^\prime}$) valley (Fig. \ref{grand}(e)). The indirect exciton has not been observed experimentally in PL, but resonant Raman measurements offers some evidence for dark excitons above \xo{}.\cite{mcdonnell_observation_2020} While the energy of this exciton matches the energy of H1, a viable mechanism is required to provide the missing momentum needed for optical recombination of this dark state. Mechanism I explores how the interface between \tise{} and \mose{} could potentially activate such dark states by introducing a new spatial periodicity that violates quasimomentum conservation.

\mose{} and \tise{} form an incommensurate superstructure for most stacking configurations, which, formally, implies a full relaxation of quasimomentum conservation. In order to illustrate this, consider the \mose{} Bloch wave functions at the \tise{} - \mose{} overlap. We assume there is no hybridization between \mose{} and \tise{} wave functions, which is reasonable due to their large spatial separation. The quasimomentum can then be defined by the \mose{} lattice alone and the Bloch functions are
$\psi_{\mathbf{k}}(\mathbf{r)=}\sum_{\mathbf{g}}a_{\mathbf{k+g}}\exp
[i(\mathbf{k+g)\cdot r]}$, where $\mathbf{g}$ is a \mose{} reciprocal lattice (r.l.) vector and $\mathbf{a_{k+g}}$ are the Bloch coefficients. The presence of the \tise{} lattice and CDW can be thought of as a periodic ``defect'' for \mose{} that enables scattering between the $\mathit{K}$ and $\mathit{Q}$ points. The matrix element of the associated potential for Bloch states is $
\left\langle \psi_{\mathbf{Q}}|V(\mathbf{r)|}\psi_{\mathbf{K}}\right\rangle$,
where $\mathbf{K}$ is a wavevector for any of the $\mathit{K}$-equivalent points and $\mathbf{Q}$ is a wavevector for any of the $\mathit{Q}$-equivalent ones (Fig. \ref{grand}(e)). $V(\mathbf{r)}$ is the effective \tise{} defect potential, which can be expanded in the \tise{} r.l. vectors $\mathbf{t.}$ Note that the lattice parameter for \tise{} is 7.8\% larger than \mose{}, so that if we measure the reciprocal space in units of $2\pi/a_{\mathrm{MoSe}_2},$ then for Mo the first r.l. vectors will be $\mathbf{g}_{1,2}=\{\sqrt{3}/2,\pm 1/2\};$ $\mathbf{g}_{3}=\mathbf{g}_{1}+\mathbf{g}_{2}.$ For \tise{}, the equivalent r.l. vectors $\mathbf{t_{1,2,3}}$ will be shorter by 7.8\%.

Using the expansion of $V(\mathbf{r})$ we obtain,
\begin{align*}
&\left\langle \psi_{\mathbf{Q}}|V(\mathbf{r)|}\psi_{\mathbf{K}}\right\rangle
 =\sum_{\mathbf{t}}\left\langle \psi_{\mathbf{Q}}|v_{\mathbf{t}%
}e^{i\mathbf{t\cdot r}}\mathbf{|}\psi_{\mathbf{K}}\right\rangle \nonumber\\
 &=\sum_{\mathbf{t,g,g}^{\prime}}\left\langle a_{\mathbf{K+g}}%
e^{i(\mathbf{K+g)\cdot r}}v_{\mathbf{t}}e^{i\mathbf{t\cdot r}}a_{\mathbf{Q+g}%
^{\prime}}^{\ast}e^{-i(\mathbf{Q+g}^{\prime}\mathbf{)\cdot r}}\right\rangle \nonumber \\
 &=\sum_{\mathbf{t,K,Q}}a_{\mathbf{K}}a_{\mathbf{Q}}^{\ast}v_{\mathbf{t}%
}\left\langle e^{i(\mathbf{K-Q+t)\cdot r}}\right\rangle \\
&=\sum_{\mathbf{t,K,Q}}a_{\mathbf{K}}a_{\mathbf{Q}}^{\ast}v_{\mathbf{t}}\delta(\mathbf{K-Q+t)}.
\end{align*}
In the last line the summation goes over all equivalent $\mathit{K}$ and $\mathit{Q}$ points. While, in principle, in an infinite lattice one can always find a triad $\mathbf{K,Q,t}$ wavevectors that closely satisfies the condition $\mathbf{|K-Q+t|=0,}$ the coefficients $v_{\mathbf{t}},$ and, to lesser extent, $a_{\mathbf{g}},$ rapidly decay (the form factor effect), and so this scattering process can only be efficient if $t$ is small. 

Determining if the \tise{} ''defect`` potential can enable $\mathit{K}$$\rightarrow$$\mathit{Q}$ scattering requires matching $\mathbf{|t|}$ and $\mathbf{|K-Q|}$. We have carried out DFT calculations of the \tise{} - \mose{} heterostructure to determine how this position is altered by interlayer coupling. The location of the $\mathit{Q}$ valley is $\kappa=|\mathbf{Q- \Gamma}|/|\mathbf{K-\Gamma}|$, which are clustered around $0.55\pm0.05$ for multiple DFT runs. The magnitude of the smallest scattering vectors $\mathbf{K-Q}$ are: $(1-\kappa)/\sqrt{3},\, (1+\kappa)/\sqrt{3},\, (\sqrt{1-\kappa+\kappa^{2}})/\sqrt{3}$, $(\sqrt{1+\kappa+\kappa^{2}})/\sqrt{3},\, (2-\kappa)/\sqrt{3}$ and $(\sqrt{4-2\kappa+\kappa^{2}})/\sqrt{3}$, where the first two values correspond to scattering from $Q$ to $K$ or $K^{\prime},$ while the next two to scattering from $Q^{\prime}$ to $K$ or $K^{\prime},$ and the last two to scattering into the next BZ. For $\kappa=0.55,$ in units of $2\pi/a_{\mathrm{MoSe}_2},$ these are 0.260, 0.895, 0.501, 0.785, 0.832, and 1.033, respectively. At the same time, the smallest vector of the r.l. of \tise{} without the CDW is $t_{1}=0.928(2\pi/a_{\mathrm{Mo}})$.

In Fig. \ref{grand}(e) we present the magnitude of the scattering wavevector $\mathbf{|K-Q|}$ between different $\mathit{K}$ and $\mathit{Q}$ points versus $\kappa$ in \mose{}. These are compared to the r.l. vectors of \tise{} in the normal (blue dots) and CDW (red dot) phases. The smallest \tise{} r.l. vector can only match one of the $|\mathbf{K-Q|}$ values if $\kappa$ is 0.607, which is far outside our DFT predictions. In the 2 $\times$ 2 $\times$ 2 CDW phase of \tise{}, the magnitude of the r.l. vectors ($\tau_i$) are shortened to : $\tau_{1}=t_{1}/2;~\tau_{2}=t_{1}\sqrt{3}/2;~\tau_{3}=t_{1};~\tau_{4}=\sqrt{7}/2t_1$. We find that particularly $\tau_{2}=0.804$ is close to the magnitude of one of the $\mathbf{K-Q}$ vectors. This is shown by the red dot in Fig. \ref{grand}(e). The mismatch is less than $0.019 (2\pi/a_{\mathrm{Mo}})$ and is reduced to zero for $\kappa=0.59,$ a value within the range of the DFT calculations.

Therefore we conclude that the CDW potential opens a new $\mathit{K}$$\rightarrow$$\mathit{Q}$ scattering channel. This would enable optical recombination of an indirect, finite momentum exciton resulting in the appearance of a new PL line at the same energy as H1.\cite{Deilmann2019} Future calculations taking into account the effect of the CDW-exciton coupling may be able to assess this scenario quantitatively, but such calculations are outside our current capabilities. This result explains the emission energy of H1, its disappearance at $T_{CDW}$, and its anticorrelation with \xt{}. However, this mechanism would also imply that H1--\xo{} (barring unrelated phenomena) to be $T$-independent and the H1/\xo{} intensity ratio to follow $\Delta^2$. Our observations indicate just the opposite (Fig. \ref{grand}(a) and \ref{grand}(c)). Furthermore, this mechanism cannot explain the decrease in the H1/\xo{} intensity ratio with excitation power in Fig. \ref{grand}(b).

\grand{}

\subsection{Mechanism II: Interlayer 2D \tise{} - \mose{} Exciton}

Another intriguing possibility is the formation of an interlayer exciton by an electron in \mose{} and a hole in \tise{} layer as shown in Fig. \ref{grand}(d). This is only possible due to the opening of the CDW band-gap in the low-temperature regime, but is inconceivable in the normal metallic phase. The intensity of such an exciton will be defined by ($T$-independent) interlayer tunneling, and the position will be, roughly, given by $E_c(\mathrm{MoSe}_2)-E_F(\mathrm{TiSe}_2)-E_{CDW}(\mathrm{TiSe}_2)/2$. The energy gap, $E_{CDW}$, is proportional to the order parameter $\Delta^2$, and its value is unclear: experiments cite different numbers, most ARPES studies find the top of the valence band to be separated from the Fermi level ($i.e.$, half the band gap) by 50--75 meV \cite{Monney2010,Chen2016b}, qualitatively consistent with the H1{}--\xo{} separation, and roughly following $\Delta^2$. 

For this scenario to be realized the top of the Mo valence band at K should fall inside the CDW gap in \tise{} (or, equivalently, within a few tens of meV from the \tise{} Fermi energy, $E_F$, in the metallic phase). Our standard DFT calculations using a supercell with 2 $\times$ 2 periodicity in \tise{} and $\sqrt{7}\times\sqrt{39}$ in \mose{} (Ti$_{16}$Mo$_{19}$) show that $E_F$ is about 230 meV above the \mose{} valence band. This is illustrated in Fig. \ref{grand}(d) and detailed in Supporting Information Section S4. So, while this naturally explains all five experimental observation, it also requires an assumption that the DFT calculations of the band alignment are off by 100--150 meV, which may be reasonable once the electronic correlations that drive CDW formation and the details of the vdW interface are properly included.

\subsection{Mechanism III: Exciton-Phonon and Exciton-Plasmon Interaction}
Exciton formation and recombination can in principle be assisted through coupling to a variety of bosonic excitations. Again focusing on the activation of a dark finite-momentum exciton, the lowest energy phonon with the appropriate momentum has an energy of $\sim 10$ meV \cite{MoSe2phonons}, so that the corresponding PL energy should be shifted down (phonon-assisted emission) from the momentum-dark exciton energy. This would suggest that H1 is a phonon replica of a higher energy exciton, either symmetry forbidden or momentum-indirect. However, this mechanism is unlikely considering that it is intrinsic to \mose{} and cannot explain the role of the \tise{} CDW. In principle, interlayer coupling could allow \tise{}  phonons to create exciton-phonon replicas in \mose{}. H1 should in this case emerge at higher temperature if \tise{} phonons are involved and the replica would show additional temperature dependence compared to \xo{}, in contradiction with our observation.\cite{Shibata1998,Huang2016,Lin2012} 
An additional possibility regarding the activation of a finite-momentum exciton by coupling to \tise{} is the presence of a distinct low-energy plasmon. A curious aspect of the \tise{}-CDW transition is the presence of a low-energy plasmon, which has been claimed as evidence of the excitonic insulator mechanism,\cite{kogar2017signatures} although this claim was later disputed \cite{lian2019charge}. This plasmon was measured to have an energy of $\hbar\omega_{pl}(q)\approx 50$ meV at $T$ = 17 K with $q=0$. This excitation was found to soften with temperature, reaching $\hbar\omega_{pl}\approx 35$ meV at $T$ = 185 K. The plasmon is only present in the CDW phase, so it is tempting to associate it with the H1 PL line.

Signatures of exciton-plasmon interaction in PL have attracted considerable attention recently \cite{VanTuan2017:PRX,VanTuan2019:PRB,Scharf2019a,VanTuan2019:arxiv}. While these papers consider excitons and plasmons spatially coexisting in the same material, the theory is equally applicable to spatially separated \mose{} excitons and \tise{} plasmons, as long as they are coupled by Coulomb interaction. In principle, two mechanisms are possible (see Supporting Information - Section S3 for a detailed discussion): the exciton Green function can be renormalized by a virtual process of emission and absorption of a plasmon or a process with either emission or absorption of a real plasmon. The former process shifts the exciton line $E_X$ up by a fixed amount, which in the first approximation can be expressed as $E_X^2-E_{X0}^2\approx4|\mathcal{M}|^2E_{X0}\hbar\omega_{pl}/(E_{X0}^2-(\hbar\omega_{pl})^2)\approx4|\mathcal{M}|^2\hbar\omega_{pl}/E_{X0}$, where $E_{X0}$ is the position of \xo{} in the absence of exciton-plasmon coupling and $\mathcal{M}$ is the exciton-plasmon coupling constant. However, one would expect the coupling constant to vary spatially, so instead of two lines one would observe a broad manifold starting at $E_{X0}$ and ending at around  $E_{X0}+2|\mathcal{M}|^2\hbar\omega_{pl}/E_{X0}$, in contradiction with the experiment. The other mechanism preserves the main line and adds two satellites, shifted down and up by around $\omega_{pl}$. The intensity of the upper peak is roughly temperature independent, and the intensity of the lower peak is proportional to the population of thermally or extrinsically excited plasmons. We do not observe the lower satellite at all, and the upper satellite, H1, only loses its intensity with temperature (Fig. \ref{temppower}(b)) and with the laser power (Fig. \ref{grand}(b)), in obvious contradiction with the assumed physics.

\section{Conclusion}

We have discovered the presence of a new exciton-like peak in \tise{} - \mose{} heterostructures using temperature-dependent PL spectroscopy. The H1 emission feature is localized to the heterostructure interface and correlated with $T_{CDW}$ in \tise{}. We have presented multiple scenarios that could explain the origin of this feature. The most plausible explanations of H1 are, presently, 1) an interlayer \tise{} - \mose{} exciton and 2) the brightening of momentum dark excitons by the CDW potential. These results are the first demonstration of exciton engineering via proximitized CDWs and provide the 2D theoretical community with a fresh challenge to understand the microscopic mechanisms underlying CDW-exciton interactions.

\section{Growth of TMD materials}
\mose{} crystals were grown by the chemical vapor transport (CVT) method using polycrystalline \mose{} powder ($\approx$ 1 g) and SeBr$_4$ transport agent ($\approx$ 0.1 g). The source and growth zones in a vacuum-sealed 20 cm long quartz ampoule were kept at 980 $\degree$C and 890 $\degree$C, respectively, for 7 days. The procedure for CVT-grown \tise{} crystals is outlined in Ref.\,\citen{Campbell2019}.

\section{Sample preparation, assembly and AFM `Nano-Squeegee'}
Bulk-hBN flakes were mechanically exfoliated and transferred on O$_2$ plasma cleaned Si/SiO$_2$ substrates. MLs of 2H-\mose{} obtained through mechanical exfoliation were subsequently transferred on an identified hBN flake. For both materials, the transfer process was done using the PDMS-based dry viscoelastic stamping method \cite{Castellanos-Gomez2014}. This method has been known to leave polymer residue between the interface of two TMDs deposited during the transfer process, but has proven to be optimal for fabricating heterostructures \cite{Frisenda2018a}. To create clean interfaces, an AFM-based `nano-squeegee' procedure was employed \cite{Rosenberger2018}. This involves the use of a standard AFM tip to push out polymer residue deposited between the two TMDs in a vertical heterostructure. We were able to use the same method to also remove surface residue present on the ML-\mose{} flake, using a 7 N/m spring constant tip and a contact force of 140 nN. Bulk \tise{} is then brought into contact with the sample using the same dry-transfer technique as done for ML-\mose. The sample was additionally vacuum annealed at 200 $\degree$C for 5 hours to improve coupling.

\section{Experimental setup}
Low-temperature PL and Raman measurements were carried out on a home-built confocal microscope setup with 532 nm laser excitation focused through a 0.42 NA, 50x long working-distance objective to achieve a spot diameter of 2.4 $\mu$m. The light is collected in a back-scattering geometry, with the collection fiber-coupled to a 500 mm focal length single spectrometer integrated with a liquid-N$_2$ cooled CCD detector. The samples were placed under vacuum and cooled in a closed cycle He-cooled cryostat (Montana Instruments Corporation) with a variable temperature range from 4 K - 300 K. Raman measurements on the \tise{} - \mose{} interface were carried out using the same collection scheme, however, the excitation path included a collection of Bragg grating notch filters to get us within 15 \cmi{} of the laser line. The excitation wavelength used for Raman measurements was 532 nm and the laser power was kept at 300 $\mu$W pre-objective. Spatially-resolved and temperature-dependent PL measurements were done using the same laser wavelength while the power was kept within 150 $\mu$W pre-objective.

\section{Acknowledgement}
P.M.V. and J.J. acknowledge support from the National Science Foundation (NSF) under Grant No. DMR-1748650 and DMR-1847782, and the George Mason University Quantum Science and Engineering Center. This work is partly supported through the Material Genome Initiative funding allocated to NIST. I.\v{Z}. was supported by the U.S. DOE, Office of Science BES, Award No. DE-SC0004890. We acknowledge valuable discussions with Madeleine Phillips and Steven Hellberg. Materials synthesis at the University of Maryland Quantum Materials Center was supported by the Gordon and Betty Moore Foundation's EPiQS Initiative through Grant No. GBMF9071.

\section{Disclaimer}
Certain commercial equipment, instruments, or materials are identified in this paper in order to specify the experimental procedure adequately. Such identification is not intended to imply recommendation or endorsement by the National Institute of Standards and Technology, nor is it intended to imply that the materials or equipment identified are necessarily the best available for the purpose. 

\section{Data availability statement}
The data that support the findings of this study are available from the corresponding author 
upon reasonable request.

%aipnum4-2.bst 2019-01-14 (MD) hand-edited version of apsrev4-1.bst

\end{document}